# Debye temperature of disordered bcc-Fe-Cr alloys

S. M. Dubiel[1*], J. Cieślak[1] and B. F. O. Costa[2]

[1] Faculty of Physics and Computer Science, AGH University of Science and Technology, 30-059 Krakow, Poland

[2] CEMDRX Department of Physics, University of Coimbra, 3000-516 Coimbra, Portugal

## Abstract

Debye temperature, $\Theta_D$, of $Fe_{100-x}Cr_x$ disordered alloys with $0 \leq x \leq 99.9$ was determined from the temperature dependence of the centre shift of $^{57}$Fe Mössbauer spectra recorded in the temperature range of 80 – 300 K. Its compositional dependence shows an interesting non-monotonous behaviour. For $0 < x \leq \sim45$ as well as for $\sim75 \leq x \leq \sim95$ the Debye temperature is enhanced relative to its value of a metallic iron, and at $x \approx 3$ there is a local maximum having a relative height of ~12% compared to a pure iron. For $\sim45 \leq x \leq \sim75$ and for $x \geq \sim95$ the Debye temperature is smaller than the one for the metallic iron, with a local minimum at $x \approx 55$ at which the relative decrease of $\Theta_D$ amounts to ~12%. The first maximum coincides quite well with that found for the spin-waves stiffness coefficient, $D_o$, while the pretty steep decrease observed for $x \geq \sim95$ which is indicative of a decoupling of the probe Fe atoms from the underlying chromium matrix is likely related to the spin-density waves which constitute the magnetic structure of chromium in that interval of composition. The harmonic force constant calculated from the Debye temperature of the least Fe-concentrated alloy ($x \geq 99.9$) amounts to only 23% of the one characteristic of a pure chromium.

[*] Corresponding author: dubiel@novell.ftj.agh.edu.pl (S. M. Dubiel)

## 1. Introduction

$Fe_{100-x}Cr_x$ alloys have been both of scientific and technological interests. The former follows, among other, from the fact that they can be regarded as a model system for studying various physical properties, e. g. magnetic ones, and testing appropriate theoretical models. The latter stems from the fact that the alloys represent a basic ingredient for a production of stainless steels that, due to their excellent properties, find a wide application in various branches of industry [1, 2]. One of the characteristic features of the Fe-Cr alloys is also that they form a solid solution within the whole concentration range while keeping the same (bcc) crystallographic structure. This is very important as it enables studying the effect of composition on various physical properties in a wide range within the same structure. In particular, as illustrated in Fig. 1, the lattice constant, for $0 \leq x \leq \sim 30$ shows some deviation from the behaviour expected from the Vegard's law, while it follows it quite well for $x \geq \sim 30$.

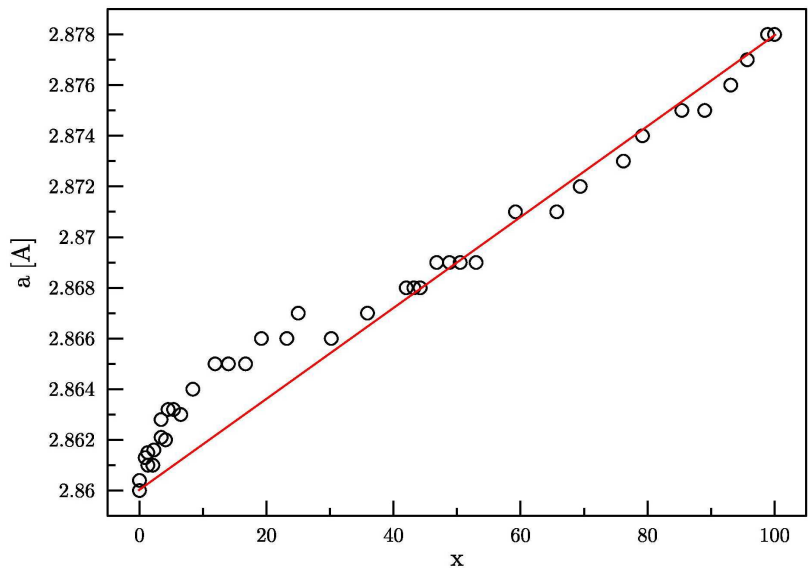

Fig.1 (Colour online) Lattice constant, *a*, versus chromium concentration, *x*, for $Fe_{100-x}Cr_x$ alloys according to [3]. The straight line stands for the behaviour following the Vegard's law.

The Curie, $T_C$, and the Néel temperature, $T_N$, are, in general, as schematically presented in Fig. 2, a monotonous function of the composition, as is the average magnetic moment, neglecting a shallow maximum observed in the Fe-rich alloys [4-6].

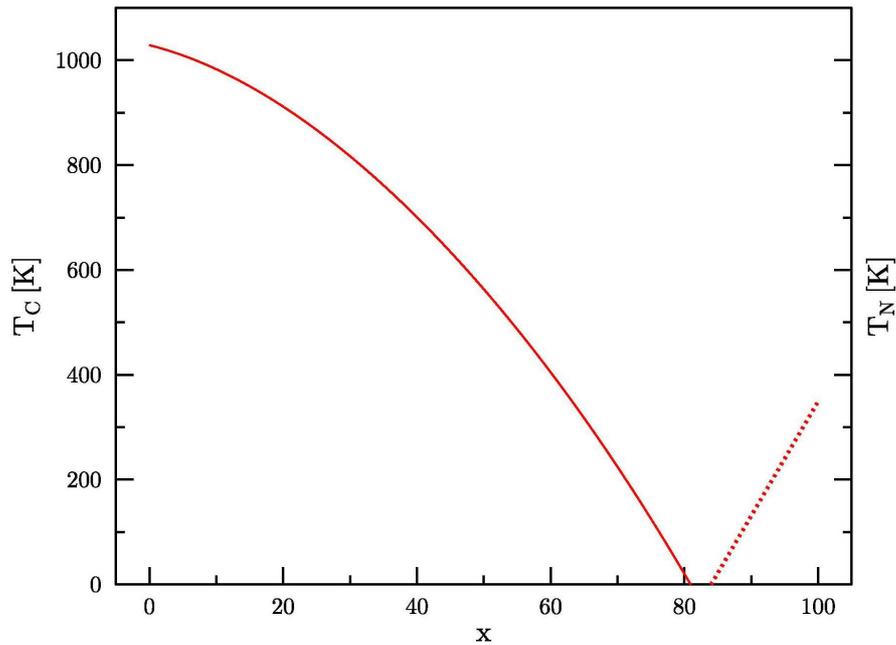

Fig. 2 (Colour online) The Curie, $T_C$, (full line and left-hand scale) and the Néel temperature, $T_N$, (dotted line and right-hand scale) versus chromium content, $x$, for $Fe_{100-x}Cr_x$ alloys. The plot has been made based on the experimental data published elsewhere [7].

On the other hand, other quantities characteristic of the Fe-Cr system, do show a non-monotonous character. As example, a module of the average $^{119}$Sn hyperfine field, $|<B>|$, as a function of chromium content, $x$, for $Fe_{100-x}Cr_x$ alloys is presented in Fig. 3. The plot has been made using the experimental data published elsewhere [8].

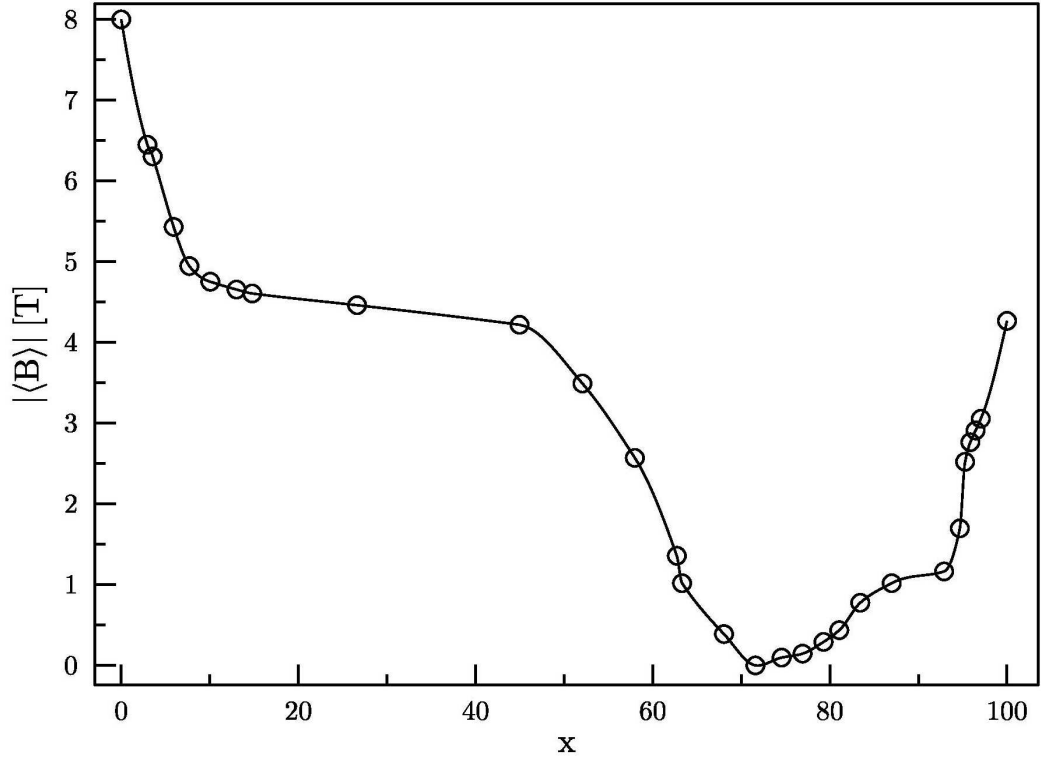

Fig. 3 The module of the average $^{119}$Sn hyperfine field, $|\langle B\rangle|$, as a function of chromium content, $x$, for $Fe_{100-x}Cr_x$ alloys, based on the experimental data published elsewhere [8].

The aim of this study was to reveal the effect of composition on the Debye temperature, $\Theta_D$, which is justified by a lack of a reliable and systematic study of this quantity in the Fe-Cr system. In addition, the Debye temperature determined from low temperature specific heat measurements for several samples showed an irregular behaviour with a difference in $\Theta_D$ as high as 374 K for $x = 80$ and 63 [9], which was not confirmed by similar measurements conducted in the temperature range of 133 – 623 K [10].

## 2. Experimental

For the present study our previously fabricated microcrystalline samples of Fe-Cr alloys [11, 12] as well as some newly prepared ones in a similar way were used. The Cr-rich alloys were made with iron in form of ~95% - enriched $^{57}$Fe isotope in order to record good quality spectra in a reasonable time. The Debye temperature was determined by means of the Mössbauer spectroscopy. For that purpose a series of Mössbauer spectra was recorded in a transmission geometry for each sample in the temperature range of 60 – 300 K using a standard spectrometer and a $^{57}$Co/Rh source of 14.4 keV gamma rays. Temperature of the samples which were kept in a cryostat was stabilized with an accuracy of ± 0.2 K. Examples of the recorded spectra both as a function of composition as well as temperature are shown in Figs. 4 and 5.

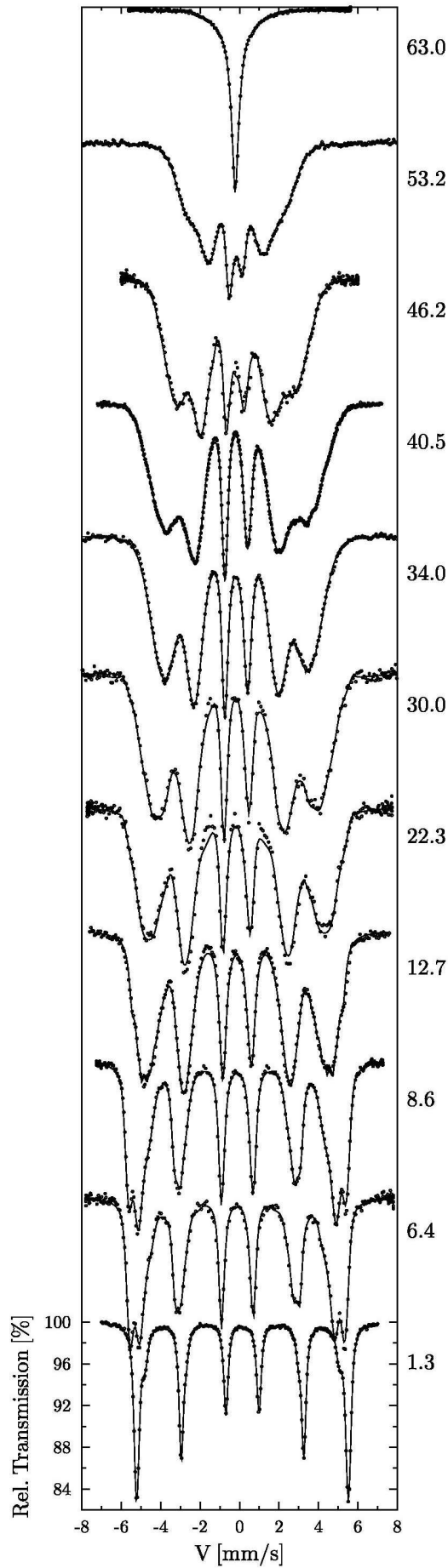

Fig. 4 Room temperature Mössbauer spectra recorded on $Fe_{100-x}Cr_x$ samples labeled with various *x*-values. The solid lines are the best-fit to the experimental data.

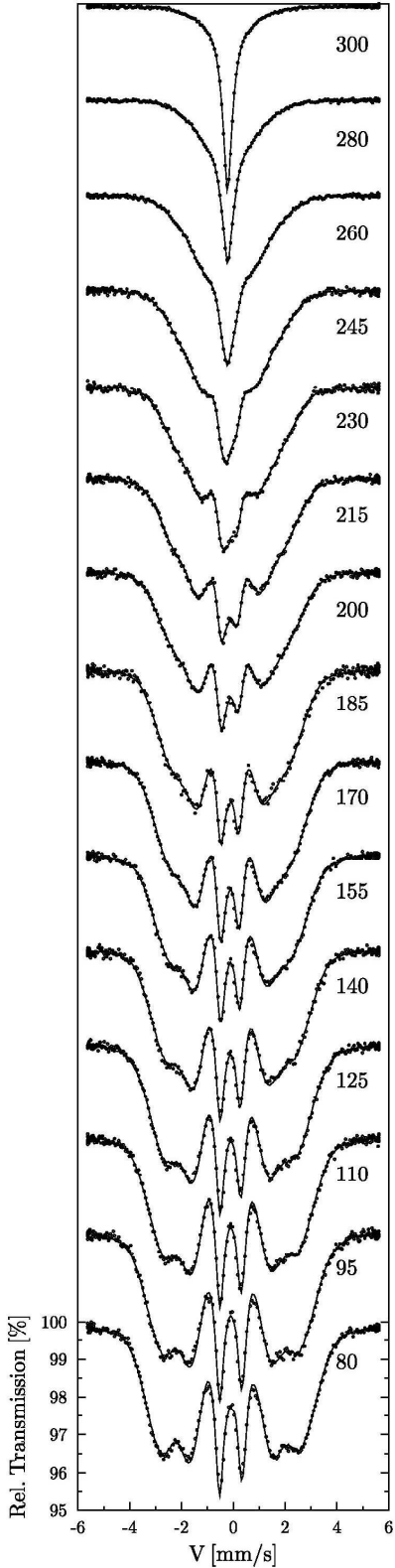

Fig. 5 Mössbauer spectra recorded on $Fe_{37}Cr_{63}$ sample at different temperatures (in Kelvin) shown. Solid lines represent the best-fits to the experimental spectra.

## 3. Results and discussion

The measured spectra were fitted to get an average value of the centre shift, *CS*, which is one of the two pertinent quantities for determining $\Theta_D$ by means of the Mössbauer spectroscopy [13,14]. The spectra with a well-resolved structure were fitted assuming that a given spectrum consists of a number of six-line pattern subspectra, each of them corresponding to a particular atomic configuration around the probe $^{57}$Fe nucleis, *(m,n)*, where *m* is a number of Cr atoms in the first neighbor shell (NN), and *n* is a number of Cr atoms in the next neighbor shell (NNN). It was further assumed that the effect of neighboring Cr atoms on spectral parameters (hyperfine field, and centre shift) was additive. Using this procedure, which has proved successful and is described in detail elsewhere [11, 12], the average centre shift, *<CS>*, could have been calculateded. The spectra with a poorly resolved structure were fitted in terms of the hyperfine field distribution method [15]. Following the experimental results [12], a linear correlation between the hyperfine field and the isomer shift was assumed in the fitting procedure. Finally, the spectra corresponding to a paramagnetic phase (Cr-rich samples) were analyzed in terms of one Lorenzian-shaped line. The Debye temperature was next evaluated from the temperature dependence of <CS> determined in the above-described ways, using the following equation:

$$<CS>(T) = IS(T) + SODS(T) \qquad (1)$$

where *IS (T)* is the isomer shift that is related to the charge density at the probe nucleus, and it has a weak temperature dependence [16], so it is usually approximated by a constant term, *IS(0)*. The latter eventually may be composition dependent. *SODS* is the so-called second-order Doppler shift which shows a strong temperature dependence. Assuming the whole temperature dependence of <CS> goes via *SODS* term and using the Debye model for the phonon spectrum one arrives at the following formula relating <CS> to $\Theta_D$:

$$<CS(T)> = IS(0) - \frac{3kT}{2Mc}\left[\frac{3\Theta_D}{8T} + 3\left(\frac{T}{\Theta_D}\right)^3 \int_0^{\Theta_D/T} \frac{x^3}{e^x - 1} dx\right] \qquad (2)$$

where *M* is the mass of the $^{57}$Fe nucleus, *k* is the Boltzmann constant, *c* is the velocity of light.

Fitting equation (2) to the *<CS>(T)* – values (whose typical behaviour is illustrated in Fig. 6), determined by the procedures described above, enabled determination of the $\Theta_D$ – values which are displayed in Table 1. In order to discuss them further and compare with the data available in the literature, we use a normalized data. For that purpose the ones found in the present study have been divided by 429 K ($\Theta_D$ = 429 K was found for a pure Fe) and those determined with another methods were divided by the corresponding $\Theta_D$-value of iron as found with a particular method (For example by 445 K in the case of the specific heat measurements). The normalized Debye temperature obtained in such way, $\Theta^*_D$, is displayed in Fig. 7 as a function of the chromium concentration, $x$, the lattice constant, $a$, and the unit cell volume, $V$.

**Table 1** Debye temperature, $\Theta_D$, and its error, $\Delta\Theta_D$, as determined in the present study for disordered bcc-$Fe_{100-x}Cr_x$ alloys. The sample of $Fe_{37}Cr_{63}$ was measured twice: in a strain-free (No 14) and in a strain (No 15) condition.

| No | $x$ [at%] | $\Theta_D$ [K] | $\Delta\Theta_D$ [K] |
|---|---|---|---|
| 1 | 0 | 426 | 14 |
| 2 | 1.3 | 457 | 11 |
| 3 | 3.2 | 482 | 13 |
| 4 | 6.4 | 452 | 9 |
| 5 | 8.6 | 439 | 11 |
| 6 | 12.7 | 471 | 16 |
| 7 | 22.3 | 463 | 29 |
| 8 | 30.0 | 433 | 25 |
| 9 | 34.0 | 473 | 15 |
| 10 | 40.5 | 447 | 13 |
| 11 | 46.2 | 412 | 16 |
| 12 | 47.8 | 411 | 29 |
| 13 | 53.15 | 380 | 24 |
| 14 | 63.0 | 397 | 19 |
| 15 | 63.0 | 419 | 15 |
| 16 | 68.0 | 423 | 44 |
| 17 | 72.8 | 377 | 24 |
| 18 | 75.0 | 427 | 13 |
| 19 | 75.8 | 413 | 17 |
| 20 | 80.0 | 472 | 14 |
| 21 | 86.7 | 470 | 20 |
| 22 | 90.75 | 479 | 16 |
| 23 | 93.0 | 482 | 26 |
| 24 | 96.0 | 507 | 18 |
| 25 | 97.0 | 470 | 7 |
| 26 | 98.0 | 474 | 11 |
| 27 | 99.0 | 467 | 18 |
| 28 | 99.9 | 395 | 14 |

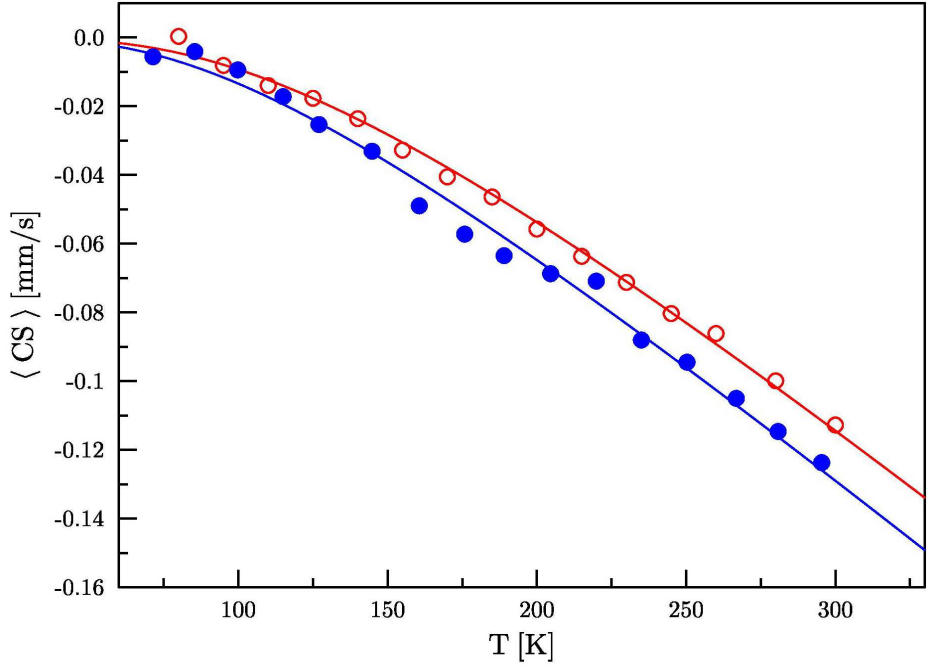

**Fig. 6**

(Colour online) Dependence of the average central shift, $<CS>$, on temperature for $Fe_{100-x}Cr_x$ alloys with $x = 53.15$ (full circles) and $x = 3.2$ (open circles). The solid line represents the best-fit to the experimental data in terms of equation (2).

Let us start the discussion with the extreme cases i.e. those for $x = 0$ and $x = 100$. They clearly illustrate the well-known fact that the value of $\Theta_D$ depends on the method applied to determine it. In case of the Mössbauer spectroscopy, there are even two ways of determining $\Theta_D$ from the same series of measurements i.e. one from the Lamb-Mössbauer factor, $f$, which is related to the square displacement of the vibrating atoms from the equilibrium position, and the other from the centre shift, $CS$, that is related to their square velocity via the second-order Doppler shift. As is evident in Fig. 7 for the pure iron, the $\Theta_D$-values obtained from the two quantities are significantly different. This, in turn, illustrates well the fact that a comparison of the Debye temperature obtained not only with different techniques but also with the same but based on different physical quantities must be done and interpreted with a caution.

Consequently, a comparison of the normalized $\Theta_D$ – values, $\Theta^*_D$, rather than the absolute ones is more justified and reasonable.

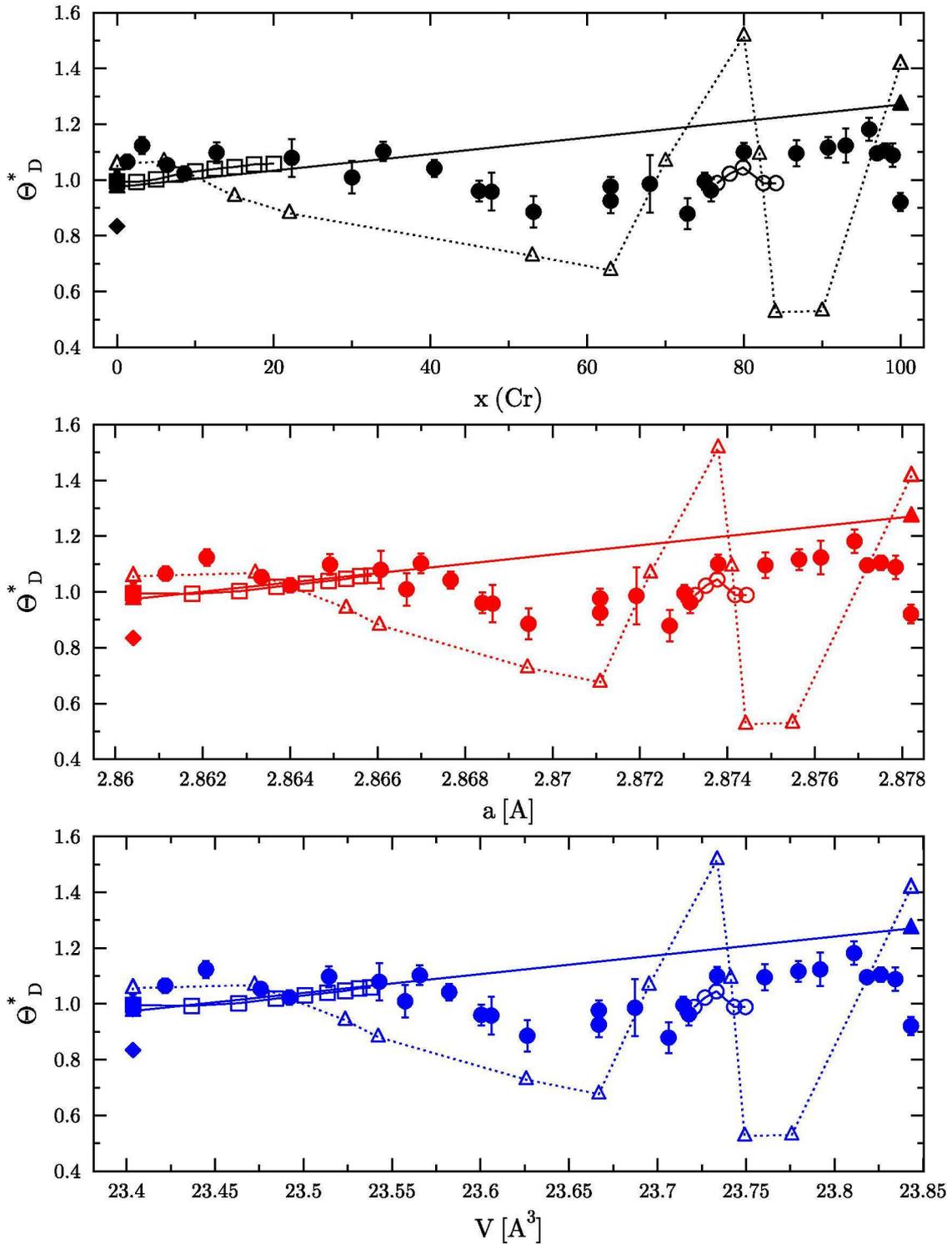

**Fig. 7** (Colour online) Normalized Debye temperature, $\Theta_D$, versus the chromium concentration, $x$ (top), the lattice constant, $a$, (middle) and the unit cell volume, $V$ (bottom).

Full circles represent the presently found data. For comparison, those obtained from the specific heat measurements are indicated by open triangles [9] and open circles [10], from XRD experiment by full triangles [17], from Lamb-Mössbauer factor by full diamond [18] and from theoretical calculations by open squares [19]. The solid line shows the behaviour expected from the Vegard´s law (for the XRD data), and the dotted one that found from the low temperature specific heat measurements [9].

It is evident that the data obtained with the present study show a complex, non-monotonous character as a function of all three parameters viz. $x$, $a$ and $V$. As there is no significant difference in the three plots displayed in Fig. 7, we will limit our discussion to the dependence of $\Theta^*_D$ (x). In general, one can distinguish concentrations for which there is an enhancement of $\Theta_D$ relative to its value for the pure iron i.e. $\Theta^*_D > 1$, and concentrations for which there is a diminution of $\Theta_D$ i.e. $\Theta^*_D < 1$. In particular, the former exist for (A) $0 < x < \sim 45$ and (B) $\sim 75 < x < \sim 95$, while the latter for (C) $\sim 45 < x < \sim 75$, and for (D) for $x > \sim 95$. The enhancement in (A), which is discussed in more detail elsewhere [20], is correlated with the behaviour of the spin-waves stiffness constant, $D_o$. The minimum in $\Theta^*_D$ observed around the equiatomic composition agrees pretty well with the concentration at which the Fe-Cr alloys are thermodynamically unstable and, on heating, they either decompose into a Fe-rich and Cr-rich phases or, for higher temperatures, they change their structure into the sigma-phase. The increase of $\Theta^*_D$ observed in (B) reflects probably the hardening effect of chromium on the lattice dynamics ($\Theta_D = 630$ K for Cr). The brake down of this trend occurring at $x \approx 95$, which obviously reflects a decoupling of the Fe atoms from the underlying chromium lattice, is very likely related to the spin-density waves that exist in the samples of these compositions. The maximum drop of $\Theta^*_D$ that happens for $x = 99.9$ reflects the weakest binding of Fe atoms to the chromium lattice. It can be expressed in terms of the harmonic force (spring) constant, $\gamma$.

There are few relevant theoretical models available [21-24]. Following Visscher's simple-impurity theory for a simple-cubic lattice [21], the quantity $\Theta_{eff}$ i.e. the effective Debye temperature as determined in the Mössbauer experiment (e. g. from equ. (2)), is related to the Debye temperature of the matrix, $\Theta_D$, by:

$$\Theta_{eff} = (M_{Cr}/M_{Fe})^{1/2}(\gamma_{Fe-Cr}/\gamma_{Cr-Cr})^{1/2}\Theta_D \qquad (3)$$

where $M_{Cr}$ and $M_{Fe}$ are the masses of the host (Cr) and the impurity atom (Fe), while $\gamma_{Fe-Cr}$ and $\gamma_{Cr-Cr}$ are the spring constants of the impurity-host and the host-host binding. $\Theta_D$ is the Debye temperature of the host (Cr). Putting $\Theta_{eff}$ = 395 K into this equation one arrives at $\gamma_{Fe-Cr}/\gamma_{Cr-Cr}$ = 0.431 which means that the coupling between Fe atoms and Cr atoms in this sample is by 57% weaker that the one between Cr atoms themselves. The reduction of the coupling is much greater if we take into account the fact that the value of $\Theta_{eff}$ = 395 K is very likely overestimated due to an unharmonic behaviour of the dynamics of Fe atoms in the temperature interval of 145 -300 K as discussed elsewhere [25, 26]. If one takes into account the value of $\Theta_{eff}$ = 292 K as determined for the harmonic mode i.e. within the temperature range of 80 – 145 K, then one arrives at $\gamma_{Fe-Cr}/\gamma_{Cr-Cr}$ = 0.235 indicating a 76.5% reduction in the spring constant value.

Gupta and Lal considered an atom undergoing an isotropic and harmonic motion and they derived the following formula for $\gamma$ [24]:

$$\gamma = \frac{Mk_B^2\Theta_D^2}{4\hbar^2} \qquad (4)$$

that can be used to determine the absolute values of the spring constant itself, if the Debye temperature is known. Using formula (4) for the present case, one obtains $\gamma_{Fe-Cr}$ = 64.8 N/m for $\Theta_{eff}$ = 395 K, and 35.4 N/m for $\Theta_{eff}$ = 292 K. The corresponding value for $\gamma_{Fe-Fe}$ = 76.4 N/m and that for $\gamma_{Cr-Cr}$ = 164.8 N/m.

Coming back to the data plotted in Fig. 7, it is obvious that our results do not confirm those found previously with the low-temperature specific heat measurements [9]. Although there are some non-monotonous changes in the Debye temperature as revealed in our study, their amplitude is not as high as reported previously, especially in the concentration range between ~80 and ~90 at% Cr [9]. On the other hand, our results agree rather well with those deduced for ~77 ≤ x ≤ ~84 from the higher temperature specific heat measurements [10] as well as with those that have been recently calculated for Fe-rich alloys [19], although the latter do not show any enhancement for $x < 5$ as found with the present investigation.

## 4. Summary

Debye temperature has been determined for disordered bcc-$Fe_{100-x}Cr_x$ alloys in the whole compositional range from the centre shift of the Mössbauer spectra recorded in the temperature interval of 60 – 300 K. The data have been corrected for the lattice constant and the unit cell volume, but this procedure has not significantly changed the character of the behaviour which has turned out to be non-monotonous. In the concentration ranges of (A) 0 to ~45 and (B) ~75 to ~95 the Debye temperature is enhanced relative to its value in the metallic iron, while in the ranges (C) of ~45 to ~75 and (D) > 95 the Debye temperature is reduced. The enhancement observed in (A) is correlated with the behaviour of the spin-waves stiffness constant, $D_o$, while the decrease observed in (D) correlates with the formation of the spin-density waves and it can be interpreted as a decoupling of Fe atoms from the chromium lattice. On the other hand, the minimum in $\Theta_D$ observed for the near equiatomic composition coincides pretty well with the composition where the alloys are thermodynamically unstable and, upon heating, they either decompose into Fe- and Cr-rich phases or they change their crystallographic structure into the sigma-phase.

**Acknowledgewment**

The results presented in this paper were partly obtained in the frame of the bilateral Polish-Portuguese project 2007/2008.